# Stumbling Through the Research Wilderness, Standard Methods to Shine Light on Electrically Conductive Nanocomposites for Future Health-Care Monitoring

Conor S Boland

School of Mathematical and Physical Sciences, University of Sussex, Brighton, BN1 9QH, United Kingdom

c.s.boland@sussex.ac.uk

**Abstract**

Electrically conductive nanocomposites are an exciting ever-expanding area of research that has yielded many new technologies for wearable health devices. Acting as strain sensing materials, they have paved the way towards real-time medical diagnostic tools that may very well lead to a golden age of healthcare. Currently, the goal in research is to create a material that simultaneously has both a large gauge factor ($G$) and sensing range. However, a weakness in the area of electromechanical research is the lack of standardisation in the reporting of the figure of merit (i.e. $G$) and the need for new metrics to give researchers a more complete view of the research landscape of resistive-type sensors. A paradigm shift in the way in which data is reported is required, to push research in the right direction and to facilitate achieving research goals. Here, we report a standardised method for reporting strain-sensing performance and the introduction of the working factor ($W$) and the Young's modulus ($Y$) of a material as two new material criteria. Using this new method, we can now for the first time define the benchmarks for an optimum sensing material ($G > 7$, $W > 1$, $Y < 300$ kPa) using limits set by standard commercial materials and the human body. Using extrapolated data from 200 publications normalised to this standard method, we can review what composite-types meet these benchmark limits, what governs composite performances, the literary trends in composites and individual nanomaterial performance and the future prospects of research.

**Introduction**

Over the past decade, nanoscience has dominated the landscape of research. From electronics[1-4], energy storage[5-8] to bioengineering[9-12], its profound influence in driving forward innovation has been felt in every corner of science. Recently, nanoscience and more specifically composites filled with nanomaterials, has drawn the attention of many researchers as tactile and health sensors due to them possessing sensitivities which far surpass that of current commercial sensing devices.[13-15] The preeminent goal of research is the creation of an 'all-in-one' sensing material that provides both high sensitivities and large working ranges.[16, 17] But, is such a material even possible? Due to a lack of performance criteria, current research in electromechanics is done so thintelligently, akin to wandering into a dark woods with no torch. With no way to see the paths taken by others in previous works due to a lack of well-defined standard procedures for reporting results, it is difficult for researchers to relate findings of others to their own or to even get an indication of how the field is progressing.[18] This leaves researchers trying to navigate the research wilderness by attempting to increase the most relatable yet least defined metric available, the gauge factor. But in doing so, we only lose sight of the purpose of our journey; the practicality of material applications.

In this review, we seek to shine a light on the need for standardisation in which data is reported in the field of electromechanics and how through this method we can create a research compass to point tech-paths in the right direction. With the reporting of the method of measuring gauge factor ($G$) varying greatly from article to article, a standardised method must be introduced to establish consistency. In considering this, when looking at the functionality of materials in application, a clear definition between strain sensing ranges and mechanical strain tolerance must also be made. To overcome this, we use a quantitative numerical value for the functional sensing range of materials, the working factor ($W$), as a new metric for defining the limitations of the electromechanical response for a given composite.

Finally, it must be noted that one of the most overlooked aspects of these nanocomposites is the importance of Young's modulus ($Y$). In order to be applied in future wearable health technologies, these materials are required to not only have large values for $G$ and $W$ but also very small values for $Y$. Using extrapolated literaturary values for $G$, $W$, and $Y$ from 200 research articles normalised to this standard procedure, we can derive information and achieve a level of understanding of electromechanical research that was previously inaccessible. For the first time we are able to identify the trends and target values in electromechanics, guiding future research out of the wilderness.

**Results**

**From Composites to Sensing Materials**

Composite materials, currently, are one of the most well studied areas in materials science research. The search for creating lighter, stronger materials is not a new revelation. In fact, it was a journey of discovery that begin thousands of years ago, with Neolithic people adding straw and wooden sticks to reinforce mud structures. With unparalleled mechanical strength, large aspect ratios and surface areas and the advances in their processability[19-22], it is to no surprise that researchers have turned towards nanomaterials, and the likes of carbon nanotubes (CNTs) and graphene, to use as filler materials to achieve such goals. With the addition of small amounts of material into a polymeric matrix we see just that, increases in the mechanical properties of the polymer.[23] As these two aforementioned materials are also electrical conductors, with the addition of further material, polymers are found to transition from insulators to conductors of electricity when a critical loading level is reached in accordance with percolation theory.[24]

Most interestingly, if one were to use a soft pliable material, such as an elastomer, stretchy nanocomposite materials with similar increases in electrical conductivity can be made.[13] Upon the application of strain, these materials display a reversible change in electrical resistance. Essentially, these materials are *strain sensors*. This change in resistance occurs due to the deformation of the conductive network with the housing matrix (Fig. 1A-B). As strain is applied to the composite, the network of conductive fillers move out of contact with one another and back again upon release. Such behaviour has been extensively investigated for CNTs and graphene-based composites[23, 25, 26] and in using that knowledge, research has expanded to demonstrate such behaviour in composites using a wide range of other nanofillers, such MoS2[27], carbon black[28], carbon fibres[29], MXenes[30], AgNWs[31] and AuNWs.[32] Typically, nano-based composite materials present values of gauge factor ($G$) ≤ 40[15], much improved over commercial metal foil strain gauges, which generally possess values between 2-7. Not often noted, nanocomposites also have an added advantage mechanically, with metal foil sensors typically made using high stiffness polymer backings that only deform to strains up to ~5% before fracture and would thus not be suitable for wearable electronics. However, with such high sensitives and flexibility, nanocomposites on the other hand have been applied as wearable bodily sensors in many forms, measuring such biomechanical functions as joint bending[33], breathing[34], pulse[35] and blood pressure.[15] Though this implies that modulus maybe an important feature of nanocomposite based electromechanical sensors, no standard has yet been set. Currently, the gauge factor is the only accepted metric for quantifying the performance of these materials. But, is this enough to determine what is a good sensor or nanomaterial or polymer matrix for strain sensing and what is not?

In fact, if we look at the use of nanomaterials in other applications, such as in Li-ion storage and clean energy harvesting, electromechanics falls flat in comparison to their regimented reporting. For battery electrode composite materials, there are a number of metrics that define performance; the conductivity, volumetric capacity, energy density, coulomb efficiency and charge/discharge times.[36] For catalyses, there is the conductivity, tafel slope, overpotential and rate of production.[37] However, in electromechanics, there is only one figure of merit but one metric is not quite enough to determine the quality of performance. By using the identification of key material properties and setting benchmarks as our light forward, we can interpret past results into a road map to help progress electromechanical research and outline what the attainable future goals are.

**Defining Performance**

As stated above, the current and only metric for which researchers can compare the performances of their strain sensing composites, is the gauge factor ($G$). Starting at low strain, resistance change can be described by

$$\frac{\Delta R}{R_0} = G \cdot \varepsilon \qquad \text{(Eq. 1)}$$

This equation describes a linear relationship between the fractional resistance change ($\Delta R/R_0$) and strain ($\varepsilon$), both dimensionless, during the deformation of the material, where *G* is the slope (Fig. 1C). This, however, in the past has been misinterpreted as the slope at any point along the response curve yielding a value for gauge factor. Many literary sources in fact quote values of *G* beginning at strains >>100%, where the matrix has yielded and the conductive network has undergone such extreme rarefication that it is artificially highly sensitive to strain due to being on the verge of failure. However, this is not the correct reporting of the metric for a material and does not allow for the reported figure of merit to be relatable to others. In terms of application, the reporting of such values is not practical, as to utilize such sensitivities it would imply that materials would need to be prestrained to 100s of percent strain. In terms of device development, having a polymeric component held at such high strain would also not be feasible due to mechanical creep.[38] Critically, Eq. 1 describes a relationship with an intercept of zero and only holds for the initial linear response of the material beginning at low strain. Points beyond this initial linear region, no longer satisfy this relationship, as fractional resistance change becomes non-linear. It is important to clarify this point, simple as it maybe, because in future applications of these composites, this proportionality will be required to calibrate electrical response to strain precisely, and vis versa.

In application, these materials are envisioned as future wearable health technologies. To function as such, they would need to measure low strain stimuli such as ones associated with pulse ($\varepsilon \sim 2\%$)[15] to strains as large as the ones associated with the bending of a joint ($\varepsilon \geq 100\%$).[39, 40] Thus, looking at our response curve and our linear fit for *G* in Fig. 1C, this would imply that a desired property of these materials would be to have a linear region or working range that extends up to $\varepsilon > 100\%$, to facilitate the measurement of both types of stimuli (i.e. an all-in-one material). With such a measurement organically arising from the fit for gauge factor, we propose that the introduction of a new metric to allow researchers to quantitatively measure the working range, the working factor (*W*), be set as a minimum standard of reporting. We define *W* simply as the strain limit (in absolute strain) at which the fractional resistance is no longer linear with strain (Fig. 1C). The introduction and adoption of such a metric would allow researchers for the first time be able to compare what filler materials and polymer systems allow for a response measurement over an extended strain range.

**Stiff Competition**

Again, keeping in mind the end goal of these materials, certain stipulations are implied. To function as skin-on sensors, the nanocomposites will be required to be very thin to accommodate adhering to the epidermis.[41] Additionally, much like the human body, the materials also are required to be soft and compliant in nature, thus having a low elastic modulus.[42] However, the mechanics of sensing materials is something that goes largely unreported in literature (see SI). Take the measurement of pulse for example (Fig. 2), this non-invasive measurement generally consisting of the sensing material being attached to the body on the wrist or neck. The artery exerts a tensile strain on the skin, which can be detected by the sensor as it deforms with the skin. As the gauge factor of our material increases, we subsequently receive more clarity in the outputted signal, with more distinguishable features of the waveform apparent. However, if a material is too stiff, this will dampen the electromechanical signal as it inhibits the deformation, regardless of gauge factor.[43] As the materials becomes stiffer and stiffer, the signal is lost in the background noise all together. As arteries only exert a small finite pressure (~5kPa)[15], there a limit to the modulus a sensing material can have that has yet to be defined.

The human body, however, has set the limits for us. Looking at the mechanical properties of skin and ligaments, respectively, they possess stiffness of ~300 kPa and ~700 kPa.[44-46] To measure unimpeded signals, the composite thus requires a stiffness comparable to that of the skin of the human body, i.e. $Y < 300$ kPa.[41, 42] In terms of functionality, as wearables, if the sensors are of a Young's modulus greater than that of the limit our bodies has set, it will plainly be too uncomfortable for the user.[47, 48] As the materials are envisioned to seamlessly be adhered to or integrated into worn clothing on the body, if they restrict motion or movement, it defeats the purpose of application. Looking back at our previous criteria, essentially, the idyllic all-in-one sensing material would be one which has $G > 7$, $W > 1$, $Y < 300$ kPa.

**Discussion**

**Gauge Factor vs Working Factor: A Negative Correlation**

Using the standardised procedure described above, values for $G$, $W$, and $Y$ were extrapolated from 200 published sources reporting the use of a nanocomposite as a sensing material, indiscriminate of filler or matrix type (see SI, Table S1). Firstly, we plot $G$ as a function of $W$ in Fig. 3A. On the same plot, the goal region, identified as a sensing material possessing $G > 7$ (the gauge factor limit of a commercial sensor) and $W > 1$ is overlaid. We set this limit for the working factor in accordance with biomechanical measurements. Joints and muscles can undergo inflections up to $\varepsilon \geq 100\%$, so a material with $W > 1$ would satisfy any measurement with an associated strain inclusive in that range. Currently, only two such materials fall within this zone, both of which utilise highly aligned CNT networks.[49, 50] For composite materials, the values for sensitivity overall fell with increasing sensing limit, from $G$ ~ 3000 at $W$ ~ 0.001 to $G$ ~ 0.7 at $W$ ~ 14. In what is an unexpected result, this decay in sensitivity in literature is found to follow a very well-defined power-law, with gauge factor universally scaling as $G \propto 1/\sqrt{W}$. Unfortunately, the underlying reasons for this inverse root dependence cannot be fully understood without a systematic study performed. However, though this is the first time such a scaling has been reported, the relation may not come as a surprise.

**Yielding Behaviours**

To understand the relationship between *G* and *W*, we must first discuss the mechanisms which controls the characteristics of the response curve and which yields a regime change from linear to non-linear behaviour. As previously stated, all composite materials, including the ones whose datasets have been sampled from for this manuscript, follow the same basic principles as described previous in Fig. 1A; the changing of inter-particle connections resulting in a change of composite electrical resistance with strain. Beginning at low strain, this resistance change is linear and dominated by conductors moving out of direct contact with one another (i.e. interfacial overlapping decreases). At a critical point, the potential barrier between adjacent materials starts to increase. Tunnelling effects then start to play a dominant role upon reaching the critical strain, with tunnelling resistance increasing exponentially with continual deformation of the composite.[51] As the average distance between filler material grows, the resistance's response to strain becomes nonlinear.[51-53] Thus, at low loading levels, though a conductive network will be more sensitive to applied strain due to fewer connections in the network, it will also lead to a lower critical limit in the linear response (i.e. working factor).[51] At the critical point, it is suggested that the strain near the tunnel gaps will be large and non-uniform so the local response may exceed the elastic linear limit at the filler interface[54, 55], corresponding to the yielding of the polymer material and an increase in the gap.[56] This implies that the working factor is proportional to that of the yield strain for a given composite system.

**Dispersal Effects**

It is well known that for composite systems, material quality and loading level can give rise to aggregation effects that effect the network structure.[57, 58] Thus, dispersion factors can result in regions of high and low conductor density which have been reported to greatly affect the linearity of a composite system's electrical response to strain.[14, 59] Poor dispersal of conductive material can lead to a scenario where network effects now dominate the linearity of the response rather than the matrix, which would lead to a less than optimum working range for a particular system. Or in other words, a working range that is lower than the maximum value set by the polymer and it's yield strain. Aggregation and dispersion factors are a common problem in most mixed phase and surface deposition systems and when they occur it can lead to the evolution of "bottle-necking" between the areas of high and low density in the network with applied strain, leading to non-linear response.[59] This can commonly be overcome though through increasing the loading levels of the filler[51, 59]; however, this is a double-edged sword. As previously mentioned, at percolation, gauge factor is at it's highest and the addition of more material to a network, decreasing its sensitivity to deformation. Conversely, though increasing the loading level would also result in an increase in the working factor, it would also increase the Young's modulus.

**Orientation Dependence**

We can use this connection between yield strain and the working factor and apply it to the Kraus model[60, 61] and from this discuss how we can exert control over the gauge factor of a system. This model considers the effect of oscillatory shear strain on a network structure and relates connection changes between fillers with the applied strain and the yield strain. The number of inter-particle connections per volume, $N$, as a function of applied strain, $\varepsilon$, can be described as:

$$N = \frac{N_0}{1+\left(\frac{\varepsilon}{\varepsilon_c}\right)^{2m}} \qquad (Eq.\ 2)$$

Where, $N_0$ is the initial number of inter-particle connections per volume and $m$ is a constant related to specific fractal dimensions of the fractal agglomerate structures and is equal to ~ 0.5. This constant is mainly a geometrical quantity of the filler network and agglomerates, independent of the specific filler or polymer type and the reasons for the universality remaining unclear.[61] $\varepsilon_c$ is also a constant related to the yield strain, or in this case the working factor, and is given by

$$W \cong \varepsilon_c = \left(k_r/k_b\right)^{\frac{1}{2m}} \qquad (Eq.\ 3)$$

Where $k_r$ and $k_b$ are the rate constants for reformation and breaking of connections. Interestingly, the Kraus model also implies that the yield strain will be constant for all loading levels of a given composite material regardless of conductor size[60, 61]. This confirms the previous assumption that the polymer's yield strain controls the maximum attainable working factor. The resistivity of a system can be related to conductor interconnectivity through a percolation-like relationship

$$\rho \propto (N - N_0)^{n_\varepsilon} \qquad (Eq.\ 4)$$

Where $n_\varepsilon$ is a percolation exponent. From our previous work[15], using Eq. 2 and 4, an expression that describes a relationship between resistivity and the yield strain (working factor) can be given by

$$\rho = \frac{\rho_0}{\left[\left(1+\left(\frac{\varepsilon}{W}\right)^{2m}\right)^{-1} + \frac{\varepsilon}{\varepsilon_t}\right]^{n_\varepsilon}} \qquad (Eq.\ 5)$$

Where $\varepsilon_t$ is a composite parameter descripted by

$$\varepsilon_t = \Lambda N_0/k_2 \qquad (Eq.\ 6)$$

And $\Lambda$ is the strain rate and $k_2$ is the rate constant for time dependent connections in the network. Eq. 5 describes both the destruction and reformation of conductor connections with strain within a composite material of very low matrix viscosity, which leads to conductor mobility. For composite materials with relatively high viscosities, leading to immobilised conductors with a reconnection constant approximately equalling zero, resistivity can be rewritten as[15]

$$\rho = \rho_0 \left(1 + \left(\frac{\varepsilon}{W}\right)^{2m}\right)^{n_\varepsilon} \quad \text{(Eq. 7)}$$

This expression implies that the application of strain reduces the number of inter-particle connections, which results in an increase in resistivity. Furthermore, this equation can be used to show that gauge factor is related to the working factor through the following relationship[15]

$$G = 2 + n_\varepsilon \left(\frac{1}{W}\right) \quad \text{(Eq. 8)}$$

Looking at Eq. 8, it implies that for large values of $G$, the working factor is intrinsically low. This describes a similar behaviour between $G$ and $W$ that was observed in Fig. 3A. Showing the validity of Eq. 7, several datasets from literature are fitted and using Eq. 8, values for $G$ derived (Fig. S1 & S2). Eq. 7 was found to fit the literary data well up to the yield strain (Fig. S1) and when comparing the calculated values using Eq. 8 to the previously measured values as a function of $W$ in Fig. S2, both are found to be in close agreement, as well as the inverse root dependence with $W$ reaffirmed. Interestingly, from Eq. 8, $G$ shows a strong dependence on the percolation exponent from Eq. 4, $n_\varepsilon$, where an increasingly larger value of $n_\varepsilon$ would yield an increase in a composite's gauge factor. If we require a sensing material with at least $W \geq 1$ and $G \geq 7$, this would mean that a value of $n_\varepsilon \geq 5$ would be required. From the fitted datasets, the calculated magnitude of $n_\varepsilon$ would seem to be dependent on the ordination of the filler. Lin et al.[62] describes a hybrid conductor system that lies in-plane with the direction of the applied strain and a calculated value of $n_\varepsilon \sim 5$. Whereas for a highly randomised system like what is reported in Wang et al[63], $n_\varepsilon$ is $\sim 0.95$. An intermediate would be Arif et al.[64], which describes a semi-structured filler system and a value of $n_\varepsilon \sim 1.7$. Essentially, in order to maximise gauge factor in a random well-dispersed system where a maximum value of working factor is achieved, one would need to improve orientation of the conductive network. However, this could prove to be a delicate balancing act as alignment of filler material would negatively impact the mechanics of the composites. In CNT nanocomposites, alignment of the fillers in the polymer matrix was found to further increase the modules by a factor of five.[22] For the purpose of the nanocomposites described here, this could be problematic if a pristine polymer has a base Young's modulus approaching that of skin, particularly as the addition of nanofillers to the matrix would push those values even closer or beyond a workable range. For future works, researchers would need to choose base elastomeric polymer matrix materials which have values of Young's modulus much lower than that of skin to allow room for mechanical variations. Additionally, alignment of fillers in a matrix can increase the electrical percolation threshold, thus requiring a higher loading level of filler material to allow for conduction of electrical current.[65] This in itself poses a problem as the mechanical properties can increase with increased loading levels or alternatively the increased loading levels could lead to higher levels of filler aggregation[22], which would negatively affect the working factor.

In general, composites from literature which control the orientation of conductive nanomaterials through lay deposition, fibre formation or microstructures, were found to have some of the higher values for $W$ whist also having relatively high values for $G$.[49, 50, 66-69] This dependency on network structure in $n_\varepsilon$ is not too dissimilar to that of the network dimensionality dependency seen for the percolation exponent in systems where percolation-like scaling in conductivity is described by the loading level of conductors in a network.[70-73] The unexpected effects network dispersion and orientation have on $G$ and $W$ explain some of the scatter in the literary data reported in Fig.3A. As the parameters are largely random throughout the data set, this would lead to values of $G$ that would be higher than expected for particular values of $W$ and vis versa.

**Gauge Factor vs Young's Modulus and the *Lay of the Land***

The relationship between $G$ and $W$ presents quite a challenge for researchers to push back against and will require a refined approach to engineering a more complete sensing material that allows both values to be maximised. However, $G$'s variation with $W$ is only one side of the story. Plotting $G$ as a function of $Y$ in Fig. 3B, again it is seen that the vast majority of datasets lie outside of the goal region, defined as $G > 7$ and $Y < 300$ kPa (the Young's modulus of skin). Unlike the plot in Fig 3A, the data falls in a cloud pattern, with no discernible trend and no intuitive link between the two metrics found. Combining the three datasets ($G$, $W$, and $Y$) into a 3-D master plot in Fig 4 however, we can truly see the lay of the land of electromechanics. From the surveyed literature, to the best of the author's knowledge, no reported sensing material possesses values that match that of the proposed optimum material. However, it would be overly critical not to note that before this review, such a large volume of data had not been laid out in such a way to see these trends or the links between these proposed sensing material parameters shown. The closest any material comes to the desired benchmark values for the optimum material is again a highly aligned/orientated CNT composite with $G = 4$, $W = 2.5$, $Y = 125$ kPa.[74]

**Conclusions**

In summary, this review has created a standardisation in the reporting of sensing material results that facilitates the communication and comparing of data and a method to which researchers can use to gauge the practicality of a material's application. Through values derived from 200 literatury sources using this method, we were able to identify the challenges for the research area; engineering materials that have both large values for $G$ and $W$ and with low values of $Y$. From this, we find that $G$ universally scales as $G \propto 1/\sqrt{W}$, inferring that sensing performances of composites maybe capped. Furthermore, we discuss how this relationship might controlled by the ordination, dispersive effects and/or the loading level of conductors in the conductive network, though further studies would be required. From this literary study, we find that to the best of the author's knowledge, no material to date meets the require benchmarks for an all-in-one material.

**Outlook**

Nevertheless, despite the previous lack of definitive performance criteria, is electromechanics stuck in a research wilderness? Well, uniquely, from using the standard method described here and the large volume of data extrapolated, interesting statistics that were previously inaccessible to researchers can be derived. Using the standardised method as a form of clarity, trends in sampled data over time in Fig. 5 from the years 2008 to 2019 show a steady overall increase in $G$ and $W$ values independently of one another (Fig. 5A & B respectively). $G$ over time is found to increase from ~14 to ~100, while $W$ on average goes from ~0.34 to ~1.2. Conversely, in Fig. 5C, an overall fall in values for $Y$ from 800 MPa to 150 MPa was observed over time. Research is instinctively going in the right direction, however, these trends in the literary data truly shows the need for the introduction of $W$ and $Y$ as new standard metrics to tie everything together when considering the data trends reported here.

Overall, the average nanocomposite strain sensor is found to have a gauge factor of $G \sim 41$ (very close to the previously assumed value), a Young's modulus of $Y \sim 300$ MPa and a working strain of $W \sim 0.62$. To compare how composites prepared using different nanomaterials perform in comparison to one another, in Fig 6, we calculate and plot the average values for $G$, $W$ and $Y$ for each nanomaterial type. In terms of materials used, graphene and graphene-like materials were the most common, followed by carbon nanotubes and multi-walled carbon nanotubes (Fig. 6A). It is to be noted, that ten materials had a count value $\leq 3$, six of which only had one count. For the forthcoming ranking of average nanomaterials parameters in literature, they are deemed to not be statistically reliable due to their low count number and are denoted by an asterisk in the following figures. In Fig. 6B, on average, hybrid-based materials (those that which utilised two or more nanomaterials) had the highest average gauge factor ($G \sim 200$). Graphene and graphene-like materials ($G \sim 50$) reported the second highest literary value, followed by CNT-based composites ($G \sim 35$). In terms of working factor (Fig. 6C), hydrogels were found to have the largest sensing ranges, at $W \sim 1.2$, which was followed by conductive polymers ($W \sim 0.6$) and CNT-based materials ($W \sim 0.55$). Hydrogels, again, were found to also have the lowest average Young's modulus ($Y \sim 200$ kPa) in Fig 6D. Amorphous carbon materials were found to have the second lowest ($Y \sim 3$ MPa) and AgNP composites the third ($Y \sim 8$ MPa). In Fig. 7, we make a 3-D master plot using the values of the three metrics ($G$ vs $W$ vs $Y$) for the individual nanomaterials. Similar to Fig 4, no nanomaterial has an average value that lies within the current goal range of the all-in-one sensing material, though hybrid-based materials would come the closest.

**Future Challenges**

The core challenge for research remains the same for the foreseeable future, i.e. to identify materials which have $G > 7$, $W > 1$, $Y < 300$ kPa. However, though the reported negative correlation between $G$ and $W$ does cast doubt on whether such an all-in-one material is achievable the modified Kraus model predicts that $G$ and $W$ can be greatly improved through engineering of the conductive filler network structure. An example of such engineering is work done by Pu et al., which demonstrated CNT/ethylene-α-octene block copolymer (OBC) nanocomposites that have a very well dispersed conductive network structure in the polymer matrix.[65] Using the author's definition of working factor and applying it to this study, it can be seen that not only does the yield strain appear to match that of the linear limit of the electromechanical response but is also constant for the composite system. Additionally, when the orientation of the CNTs was altered, so that all the fillers were aligned in one direction, the gauge factor dramatically increased. The model here wholly predicts the electromechanical behaviour of the nanocomposites in this study and proves that utilising it may in fact be an intuitive way forward in improving and understanding performances. Interesting, this model also uniquely gives researchers the opportunity to possibly revisit nanocomposite materials that had exceptional properties in the past and further optimise their performance through the better sense of perspective provided by this review.

Playing devil's advocate, perhaps the argument of all-in-one materials is a red herring, as plenty of sensing materials can measure small stimuli and others large. This problem of measuring both in the future could simply come down to using two types of sensing materials in conjunction with one another. A very much achievable goal for current research would be the study of signal hysteresis and conditioning, both largely unreported phenomena but equally important for commercialisation. A good sensing material would need to report a response that is undiminished with cycle number but any links between materials properties and low hysteresis has yet to be drawn. Regardless of goal, this review gives us collectively as researchers the ability to take previous studies and use them as a springboard to progress the field. Only through the understanding of the limitations of our past work can we move the field forward with perspective, towards real application potential, towards the forest edge.

**Methods**

Data from literaturary sources was extracted and fitted using the "Digitizer" function in Origin 2019 software. As calculated values are extrapolated from published and not raw data, all values are to be taken as an estimate. Only sources that reported or presented data that allowed for values of gauge factor, working factor and Young's modulus to be calculated we used in this review.

For further information on the derivation of the modified Kraus model see Ref 15.

**Figures:**

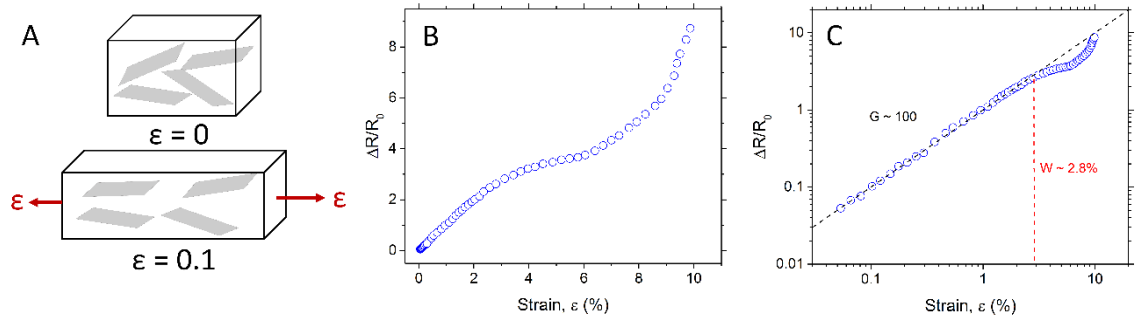

**Figure 1:** A) Top, polymer matrix with embedded electrically conductive nanosheet network in its initial state. Bottom, matrix in its strained state resulting in rarefication of the network. B) Typical electromechanical response curve of nanocomposite material being deformed from its initial state ($\varepsilon=0$) to a strained state ($\varepsilon=0.1$). C) Log-log plot of Fig. 1B with linear region fitted with Eq. 1 to extrapolate gauge factor, $G=100$. The limit to the linear region, and thus its sensing range, is defined as 0.028%, the working factor ($W$).

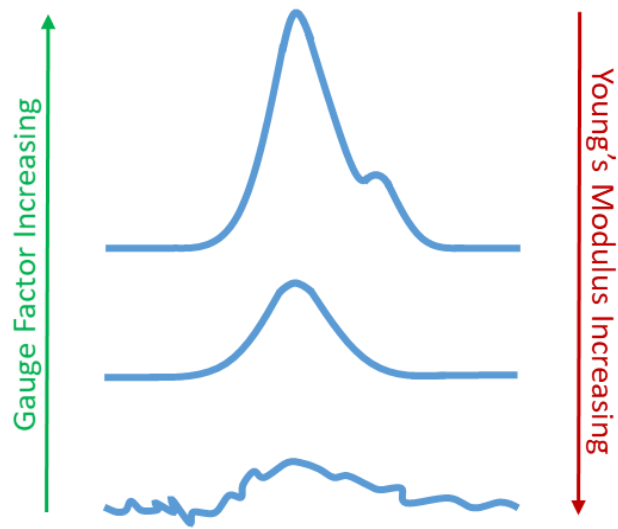

**Figure 2:** Effect of increasing gauge factor and Young's modulus on the response curve of a pulse measurement.

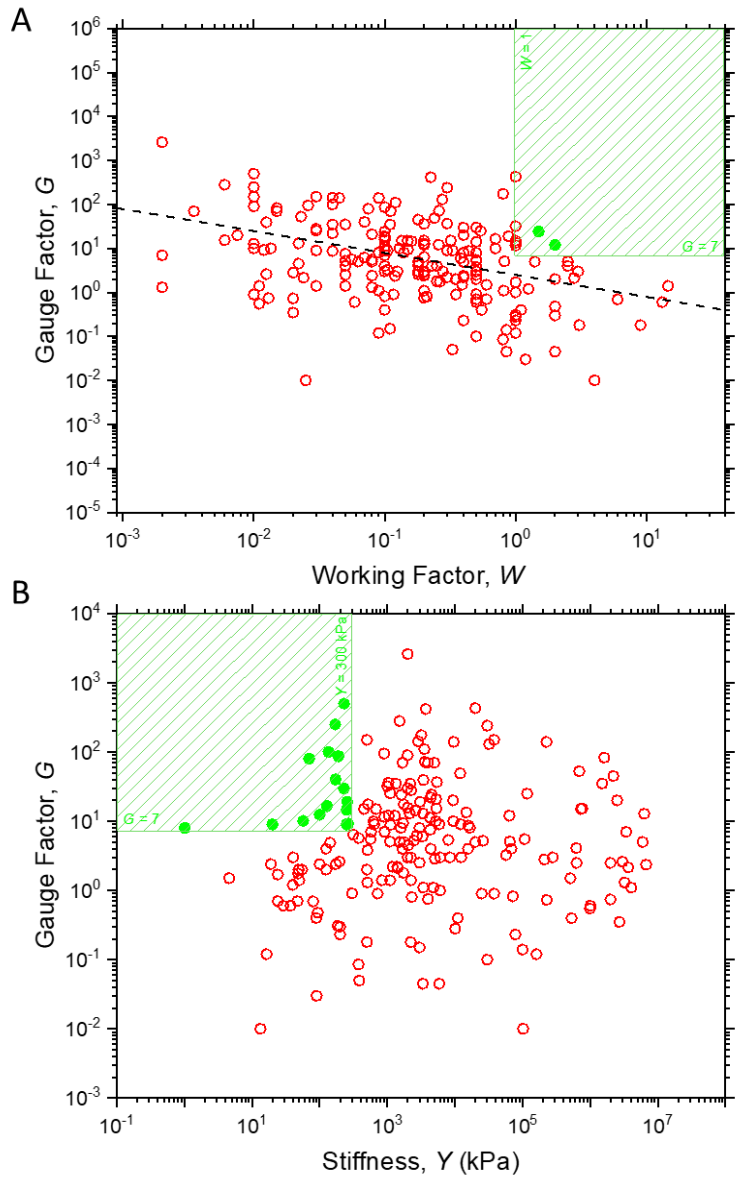

**Figure 3:** Gauge factor (G) as a function of working factor (*W*) and Young's modulus (*Y*). (A) The target region is designated by the values of *G* > 7 and *W* > 1. The dashed line represents an apparent intrinsic decay in the gauge factor of nano-based strain sensors with the working factor described by $G \propto 1/\sqrt{W}$. (B) The target region is designated by the values of *G* > 7 and *Y* < 300 kPa.

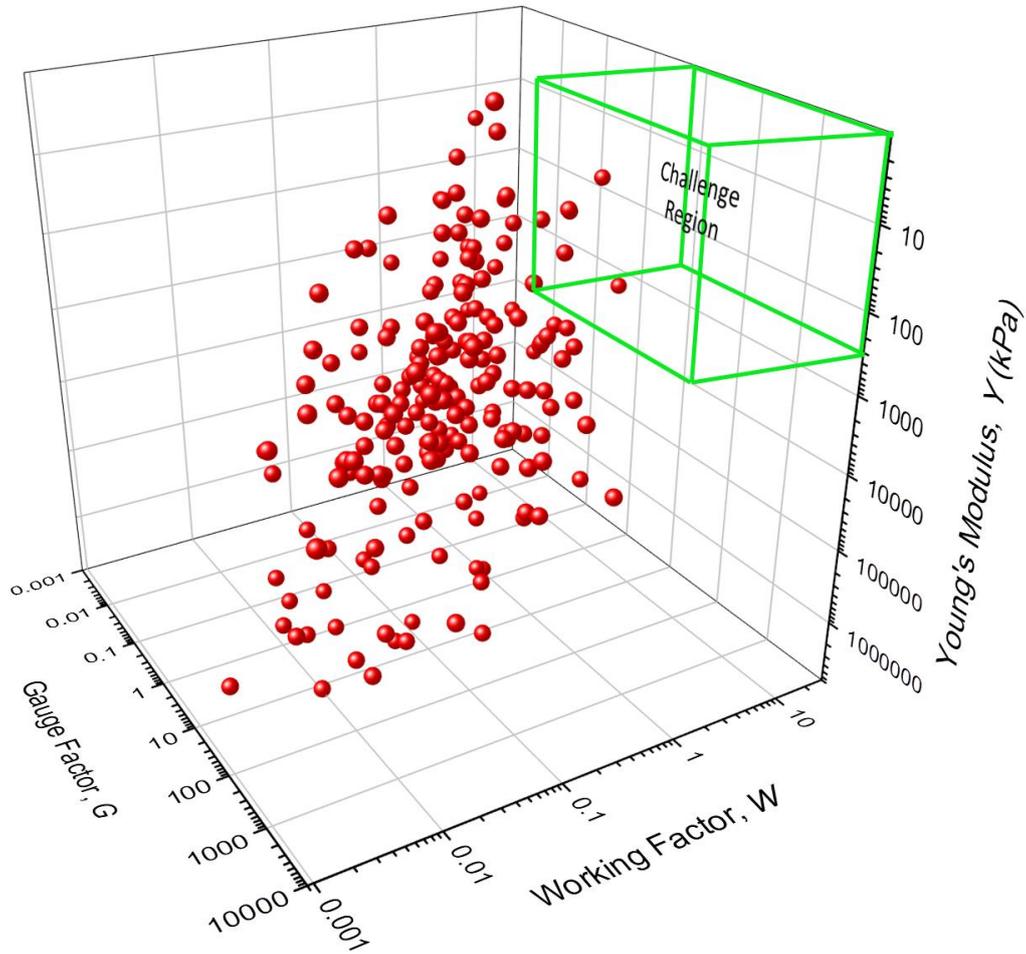

**Figure 4:** Plot of extrapolated literary values for gauge factor (*G*) vs working factor (*W*) vs Young's modulus (*Y*). The challenge region is designated by values of *G* = 7, *W* = 1, and *Y* = 300 kPa.

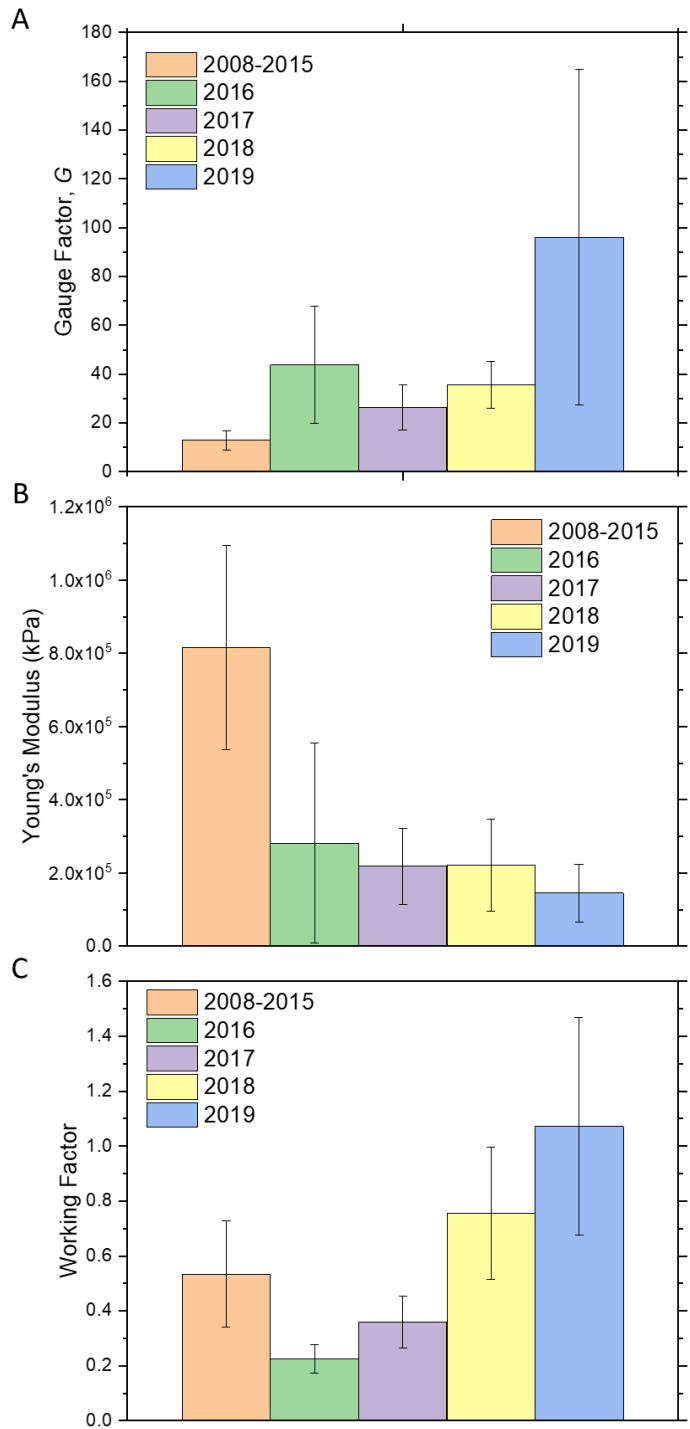

**Figure 5:** Evolution of gauge factor (*G*), Young's modulus (*Y*) and working factor (*W*) over time for sampled literature.

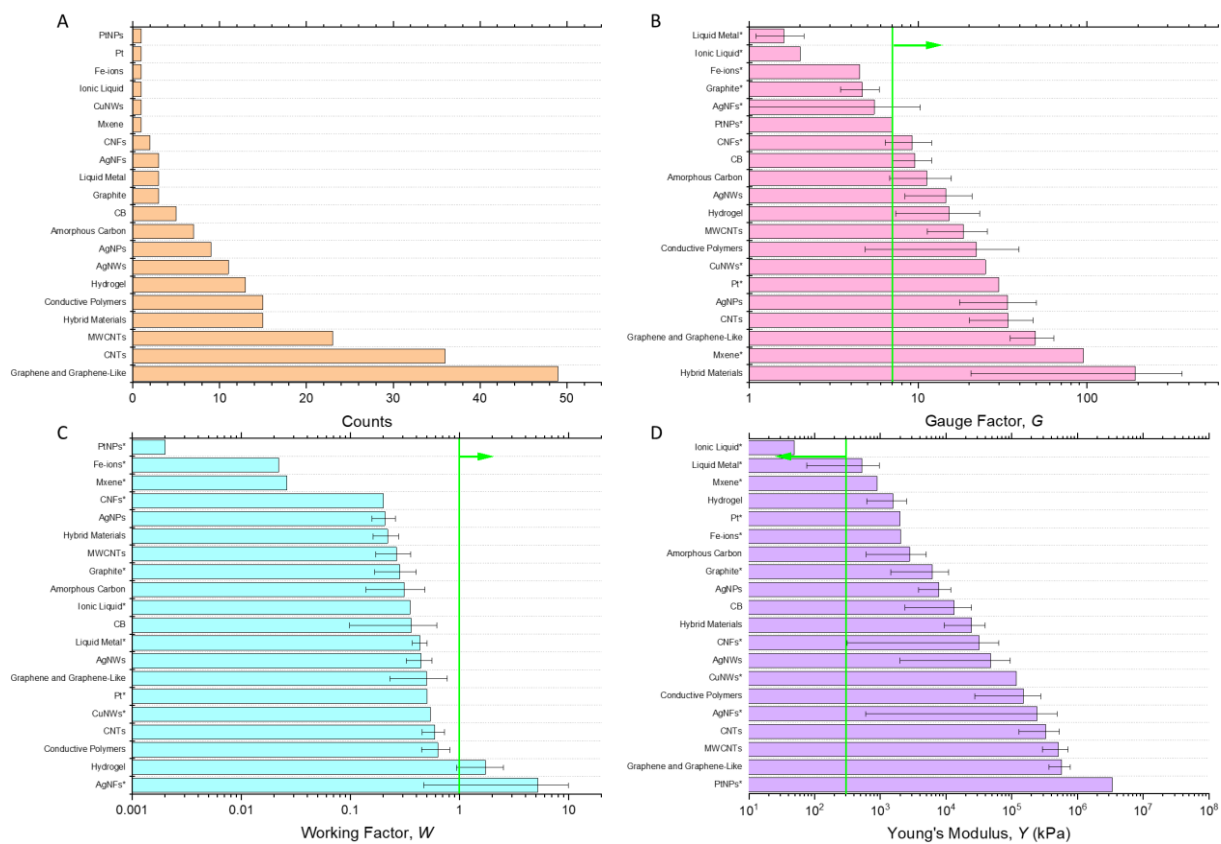

**Figure 6:** (A) Distribution of nanomaterials used in sampled literary sources. (B-C) Average gauge factor (*G*), working factor (*W*) and Young's modulus (*Y*) of materials samples from literature. Green lines and arrows denote the optimum value for each parameter and to which direction values need to tread towards to achieve surpassing it. Asterisk denotes materials with a count number ≤ 3.

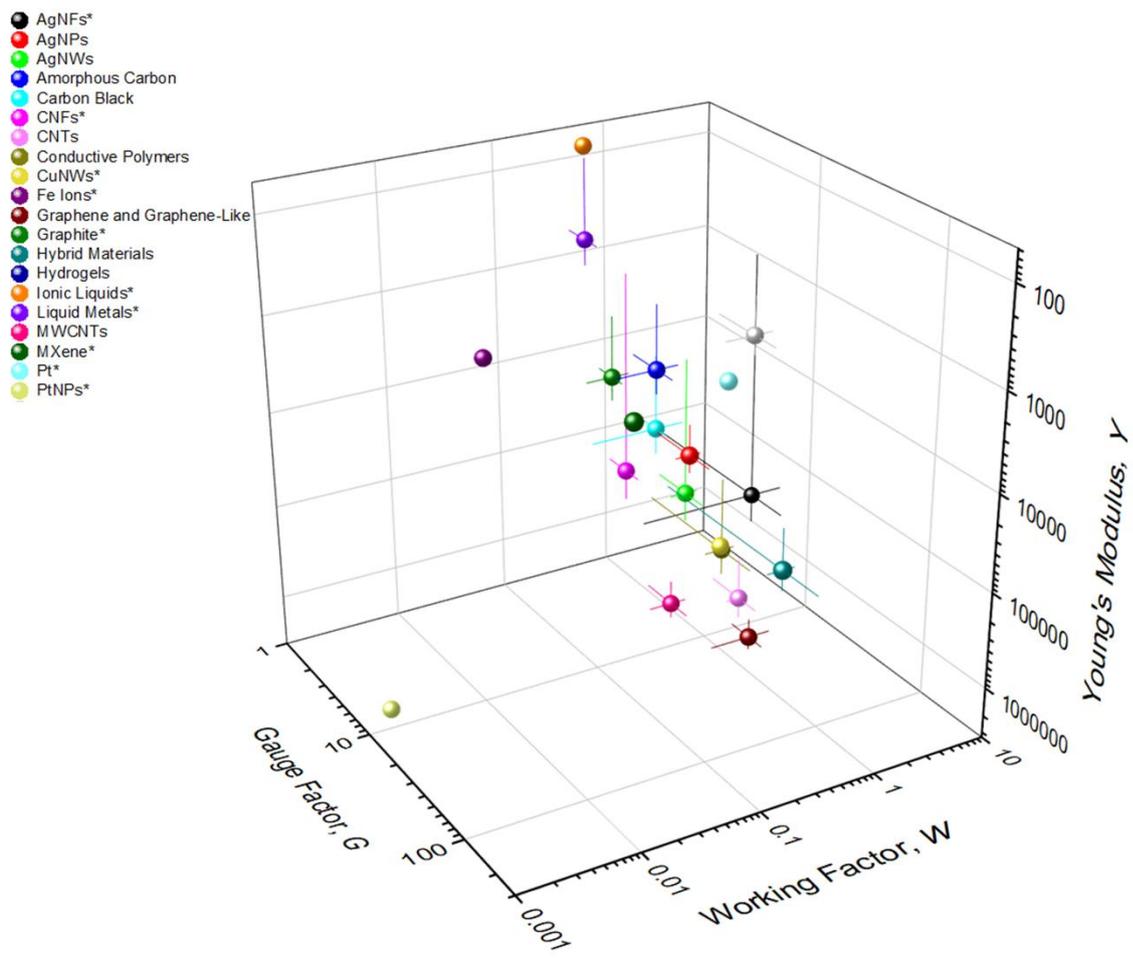

**Figure 7:** Plot of extrapolated literary values for gauge factor (*G*) vs working factor (*W*) vs Young's modulus (*Y*) of each material type sampled. Asterisks denotes materials with a count number ≤ 3.